\documentclass[prd,showpacs,superscriptaddress,nofootinbib,floatfix,preprint,11pt]{revtex4}
\usepackage{amsmath,amsfonts,mathrsfs,array, graphicx}
\usepackage[caption=false]{subfig}
\newcommand{\mc}[1]{\mathcal{#1}}

\newcommand{\ul}[1]{\underline{#1}}
\newcommand{\Oslash}{\mc{O} \hspace{-1.4ex}/\hspace{.5ex}}
\newcommand{\Lslash}{\mbox{\it\L}}

\newcommand{\T}{\text{T}}
\newcommand{\HB}{\hbar_*}  
\newcommand{\rmi}{\text{\rm{i}}}
\newcommand{\q}{\text{\rm{q}}}

\begin{document}

\title{Quantum gravity and the cosmological constant: lessons from two-dimensional dilaton gravity}

\author{Jan Govaerts} \email{jan.govaerts@uclouvain.be}
\affiliation{Centre for Cosmology, Particle Physics and Phenomenology (CP3), Institut de Recherche en Math\'ematique et Physique, Universit\'e catholique de Louvain, Chemin du Cyclotron 2, B-1348, Louvain-la-Neuve, Belgium}
\affiliation{International Chair in Mathematical Physics and Applications (ICMPA-UNESCO Chair),
University of Abomey-Calavi, 072 B. P. 50, Cotonou, Republic of Benin}
\author{Simone Zonetti} \email{simone.zonetti@uclouvain.be}
\affiliation{Centre for Cosmology, Particle Physics and Phenomenology (CP3), Institut de Recherche en Math\'ematique et Physique, Universit\'e catholique de Louvain, Chemin du Cyclotron 2, B-1348, Louvain-la-Neuve, Belgium}

\date{\today}


\begin{abstract}
In the investigation and resolution of the cosmological constant problem the inclusion of the dynamics of  quantum gravity can be a crucial step. In this work we suggest that the quantum constraints in a canonical theory of gravity can provide a way of addressing the issue: we consider the case of two-dimensional quantum dilaton gravity non-minimally coupled to a $U(1)$ gauge field, in the presence of an arbitrary number of massless scalar matter fields, intended also as an effective description of highly symmetrical higher-dimensional models. We are able to quantize the system non-perturbatively and obtain an expression for the cosmological constant $\Lambda$ in terms of the quantum physical states, in a generalization of the usual Quantum Field Theory approach. We discuss the role of the classical and quantum gravitational contributions to $\Lambda$ and present a partial spectrum of values for it.\end{abstract}
  
\pacs{04.60.-m, 04.60.Kz, 98.80.Jk,95.36.+x}

\maketitle

\tableofcontents
\section{Introduction}
\label{sec:Intro}
The cosmological constant problem\footnote{In the literature also dubbed \emph{old} in distinction from the \emph{new} cosmological constant problem, or \emph{coincidence} problem.} (see \emph{e.g.} \cite{Carroll:2000fy,Bousso:2012dk} and references therein), in its striking simplicity, is perhaps the strongest evidence that the present understanding of everything we generally refer to as ``quantum gravity'' is far from being satisfactory: while we know with a great accuracy that $\Lambda$ is very small in Planck units ($\Lambda \sim 10^{-123}$) \cite{Perlmutter:1998np,Riess:1998cb}, there is no theoretical approach that seems to be able to predict such a measurement in a convincing fashion. The most direct attempt to a solution, in which the cosmological constant is calculated in the semiclassical approximation as the vacuum expectation value of the stress energy tensor of the Standard Model of particle physics, turns out to be catastrophic, with a discrepancy between the theoretical and experimental value of $\Lambda$ between $60$ and $120$ orders of magnitude.\\
It is common understanding that the cosmological constant is related to vacuum energy: at the classical level this relation is trivial and it simply consists in noticing that in Einstein's equations the cosmological constant term can be seen as a generated by the matter sector. One could then consider the possibility of a bare cosmological constant (an actual fundamental parameter of the theory) that counterbalances all the contributions of the classical matter sectors. While this works out fine in the classical case, in the quantum theory one has to deal with the quantum fluctuations of the vacuum: the vacuum expectation value of the stress-energy tensor in the framework of quantum field theory is given by different contributions proportional to the (fourth power of the) different cutoff scales of the different sectors of the Standard Model. At this point we have already $\Lambda \sim 10^{-60}$, which rise to $\Lambda \sim 1$ if we push cutoffs up to the Planck scale, assuming no new physics. An extreme fine tuning would then be required for all those contributions to cancel out up to $123$ decimal places\footnote{We will refer to this as the ``QFT approach'' throughout this paper.}.\\
What can be done to improve this picture? The main concern with the standard QFT analysis of the vacuum energy, in our opinion, is represented by the semi-classical framework. The vacuum energy is calculated on a flat background geometry, discarding completely any contribution which could be given by the gravitational sector. While it is true that the vacuum fluctuations of a quantum gravitational field are suppressed by the Planck scale with respect to (w.r.t.) the vacuum fluctuations of the matter fields, the dynamics of quantum gravity is completely ignored, discarding the possibility of some mechanism induced by quantum gravity that might explain the value of $\Lambda$ that we measure today.\\
The aim of this work is to explore how the cosmological constant can be determined in a quantum theory of gravity, accounting for quantum dynamics of matter, gauge and the gravitational fields, including their vacuum fluctuations, possibly improving our understanding of what $\Lambda$ is really made of. After sketching the general idea we will focus on the case of two-dimensional dilaton gravity. Most of the technical results are contained in \cite{Govaerts:2011p3916,Zonetti:2011ky,Zonetti:2012fk}.\smallbreak
This paper is organized as follows: in Section \ref{sec:ccinqg} we present a general mechanism that can determine the cosmological constant in a quantum theory of gravity. Section \ref{sec:ccin2d} introduces two-dimensional dilaton gravity, its dual action in terms of Liouville fields and the inclusion of scalar matter. The BRST (Becchi-Rouet-Stora-Tyutin) formulation of the model is briefly discussed. In Section \ref{sec:quantization} we deal with the quantization of the theory, determining the central extensions of the Virasoro algebra and discussing the realization of the quantum constraints. We analyze the contributions to the value of the cosmological constant and and obtain a partial spectrum for it. We summarize and discuss our results in Section \ref{sec:discussion}.

\section{The cosmological constant in a quantum theory of gravity}
\label{sec:ccinqg}
Two interesting examples of a cosmological constant determined by the symmetries in quantum gravity are given by one-dimensional gravity \cite{Govaerts:2004ba} and two-dimensional Liouville gravity \cite{Govaerts:2011p3916}, in the presence of scalar matter. In both cases the presence of diffeomorphisms constraints was exploited to obtain a set of equations that determine the value of the cosmological constant in terms of the physical quantum states. We can generalize the principles contained in these works to the case of quantum gravity in any number of space-time dimensions $d$.\\
In a completely general way diffeomorphism invariance, in the Hamiltonian formulation, provides classical constraint equations on phase space, some of which will also include the cosmological constant:
\begin{equation}\label{classc}
\mc{H}^\mu \left (\Lambda, \dots\right )=0 \ ,
\end{equation}
where the dots indicate dependence on any additional fields included in the theory. Additional gauge symmetries provide additional constraints, so that $\mu=1,\dots,d+N$, $d$ being the number of space-time dimensions and $N$ the number of additional gauge symmetries. In order to build a consistent quantum theory we can apply the standard Dirac's approach to constrained dynamics. In particular this procedure consists in performing, in this order, gauge fixing, quantization and reduction (enforcement of the quantum constraints)\footnote{Let us remark that this is fundamentally different from gauge fixing, reduction and then quantization. This non-commutativity of quantization and reduction is thoroughly described, for instance, in \cite{Govaerts:1989xj,Govaerts:1991,Ordonez:1995kz}. An explicit example is also given in \cite{Govaerts:2004ba} in the case of 0+1d gravity with a cosmological constant coupled to a harmonic oscillator. The physical Hilbert spaces differ by an infinite number of states between the quantization-reduction procedure and reduction-quantization procedure. In particular it is shown that reduced phase space quantization is not able to reproduce the correct quantum physical content.}. When \eqref{classc} are turned into quantum constraint operators the equations become conditions on the physical states of the model:
\begin{equation}\label{genqc}
\hat{\mc{H}}^\mu (\Lambda) | \psi_{phys} \rangle=0 \ ,
\end{equation}
in a way dependent on the value of the cosmological constant and spatial coordinates, through the dependence of fields and momenta on them. This same condition can be seen as a way to determine the value of the cosmological constant required for a given quantum state to be physical.
To this purpose let us recall that it is natural for physical gauge invariant states to carry a dependence on the values of the free parameters. This can be seen in two simple examples discussed in detail in \cite{Govaerts:2004ba}. If we consider the case of 0+1 gravity with a cosmological constant coupled with a generic matter sector it can be easily shown that the cosmological constant value must coincide with the energy of the solution at the classical level. In the quantum system the combination of Schr\"odinger equation and quantum constraints determines that for the set of quantum physical states not to be void the cosmological constant must belong to the spectrum of the Hamiltonian of the matter sector, consistently with the classical result. This in turn is consistent with the time-independence of quantum physical states dictated by time-reparametrization invariance.\\
For a second example we can look at the BRST quantization of a harmonic oscillator coupled once again to 0+1d gravity with a cosmological constant: the requirement of physical states to be annihilated by the BRST charge and to be of minimal ghost number also in this case forces the cosmological constant to belong to the matter energy spectrum.\\
We can see that in both examples the choice of a specific value of the cosmological constant determines different physical Hilbert spaces and, vice-versa, the choice of a specific quantum physical state requires a specific value of the cosmological constant. This fact is simply a direct consequence of a correct quantization of a system with gauge constraints, especially in the case of diffeomorphism gauge invariance, and fits directly within Dirac's approach to the quantization of constrained dynamics.\\
Therefore, going back to the general case of \eqref{genqc}, once a basis is chosen in Hilbert space one can solve the set of equations:
\begin{equation}\label{cceq}
\langle  \psi_{phys} | \hat{\mc{H}}^\mu (\Lambda) | \psi_{phys} \rangle=0
\end{equation}
in the parameter space spanned by the the cosmological constant itself, the set of parameters which label quantum states, as well as any undetermined coupling constants. We can claim that \eqref{cceq}, independently from the number of space-time dimensions and as long as diffeomorphism invariance holds,  provides a correspondence between quantum states and the value of $\Lambda$. To turn the formal equations \eqref{cceq} into something able to provide an actual result for $\Lambda$ it is of course necessary to have fully quantized the theory, so that the explicit form and algebra of $\hat{\mc{H}}^\mu$ are known.
\section{2d Dilaton Maxwell gravity coupled with scalar matter}
\label{sec:ccin2d}
In a generalization of the one-dimensional and the two-dimensional Liouville gravity cases  \cite{Govaerts:2004ba,Govaerts:2011p3916} it is interesting to look at the case of two-dimensional dilaton gravity. Two-dimensional generalized dilaton theories provide a very interesting class of models, widely studied in the last two decades. They proved to be very useful in the understanding of classical and quantum gravity, allowing to face conceptual issues also relevant to higher dimensions. This is mainly because by eliminating part of the higher dimensional structure they benefit from significant technical simplifications, allowing for example fully non-perturbative results. An abundance of models has been studied, for example describing black hole thermodynamics and Hawking radiation and obtaining a full non-perturbative quantization of geometry in the path integral approach, with ``virtual'' black hole states in the scattering of matter fields. 
Most of the results are well summarized in \cite{Grumiller:2002nm,Grumiller:2006rc,Grumiller:2007ju}.\\
We consider the most general action principle of two-dimensional dilaton gravity in the presence of a Maxwell field:
\begin{equation}\label{dil-max_action}
 S_{DM}=\int_\mc{M} dx^{2}\sqrt{-g}\left(XR-U(X)X_{,\mu}X^{,\mu}-2V(X)-\frac{1}{4}G(X)F_{\mu\nu}F^{\mu\nu}\right)\ ,
\end{equation}
where $X$ is the dilaton, $U(X)$, $V(X)$ and $G(X)$ are arbitrary functions of $X$, and $F_{\mu \nu}$ is the usual field strength for the vector gauge field $A_\mu$. Commas denote derivatives. In \cite{Zonetti:2011ky} it was realized that, with specific restrictions on the $U$ and $G$ potentials, there exists a dual formulation of \eqref{dil-max_action} in terms of two decoupled Liouville fields on flat space-time. In particular the dual action principle takes the form:
\begin{equation}\label{Ldual}
S_{dual}=\int_\mc{M} d^{2}x \sqrt{-g_{\flat}} \left [\frac{1}{2}\left(Z_{,\mu}Z^{,\mu}-Y_{,\mu}Y^{,\mu}\right)-2\Lambda e^{Z/\xi}-\frac{e^{-Y/\xi}}{16 \q \Lambda}F_{\mu\nu}F^{\mu\nu}+\xi\left(Z-Y\right)R_{\flat}\right ] \ ,
\end{equation}
where $Z$ and $Y$ are two real scalar fields, $\Lambda$ is the cosmological constant and $\q$ and $\xi$ are non-vanishing (coupling) constants. The line element reads:
\begin{equation}
dx^{2}=-\lambda_{0}\lambda_{1}dt^{2}+(\lambda_{0}-\lambda_{1})dt\ ds+ds^{2}\ ,
\end{equation}
which is pure gauge. The subscript $\flat$ indicates gravitational quantities that are calculated w.r.t. this metric. In this formulation the cosmological constant can always be reabsorbed by a constant shift of $Z,Y$, as in $Z \to Z-\xi \ln \Lambda$, but it is left explicit in the following.\\
In the general case the usual Gibbons-Hawking-York boundary term has to be included to properly define the action at the boundary. In dilaton gravity this term has the form:
\begin{equation}
S_{GHY}=-\frac{1}{2}\int_{\partial \mc{M}}dx\sqrt{\gamma}XK \ ,
\end{equation}
where $K$ is the extrinsic curvature, $\gamma_{\mu\nu}$ is the induced metric on the boundary and $\gamma$ is its determinant. In the dual formulation one has:
\begin{equation}
S_{GHY}=-\int_{\partial \mc{M}}dx\ (Z-Y)\left(\xi K_{\flat}+\xi^{2}\left(1+8g_{\flat}\right)\partial_{t}Z\right)\ .
\end{equation}
Note how the cosmological constant does not appear in this expression, as a shift in both $Z,Y$ is irrelevant. The inclusion of massless scalar matter in the dual formulation is trivial and it is simply done by adding an arbitrary number $N$ of terms in the form:
\begin{equation}\label{Lscalar}
\mathcal{S}_\phi = -\frac{1}{2} \int_\mc{M} d^2x \sqrt{-g_{\flat}}\phi_{,\mu}\phi^{,\mu}\ .
\end{equation}
The existence of this duality is of fundamental importance for the successful application of the ideas presented in Section \ref{sec:ccinqg}: a highly coupled non-linear system is cast into a much simpler decoupled one, opening the way for canonical non-perturbative quantization. This comes at the fair price of a restriction on the $U,G$ potentials. In particular one requires:
\begin{subequations}\label{dconditions}\begin{align}
&U(X)=\frac{1}{2\xi^2}-\partial_{X}\ln(V(X)) \ , \label{Ucondition}\\
&G(X)=\frac{\xi^2 e^{X/\xi^2}}{4 \q V(X)} \ . \label{Gcondition}
\end{align}\end{subequations}
Furthermore we also ask $V(X)\neq0$. Such a restriction allows still for enough freedom to cover a large class of dilaton gravity models, for instance a subset of the so-called \textit{ab-}family (including the Witten black hole and the CGHS (Callan-Gidding-Harvey-Strominger) models \cite{Witten:1991yr, Callan:1992zr}) and Liouville gravity \cite{Nakayama:2004vk}. \smallbreak
In the following we will focus on a model described by \eqref{Ldual} with the addition of $N$ massless scalar fields \eqref{Lscalar}. In order to take advantage of the formal equivalence of field modes and quantum harmonic oscillators, which will allow to express quantum field operators in terms of creation and annihilation operators on a suitably defined Fock space of quantum states, we will restrict our attention to the case of a space-time with a cylindrical topology. In particular we will consider $\mc{M} = \mathbb{R} \otimes S^1$, where the time coordinate $t$ takes values on the real line and the space coordinate $s$ is replaced by an angular coordinate limited to the interval $[0,2\pi)$. In this process we are implicitly introducing a length scale $\ell_c$ characteristic of the size of the compactification. In the following we will work in units which give $\ell_c=1$. We are entitled to this choice in the two-dimensional case: the \emph{natural} units are usually chosen by fixing $c=\hbar=G=1$. In two dimensions, however, Newton's constant is dimensionless, so that instead of fixing $G$ we can fix the compactification scale $\ell_c$. Therefore we can simply replace the spatial coordinate $s \in \mathbb{R}$ with $s\in [0,2\pi)$ with no ambiguities.\\
All fields will be then required to be periodic in the space coordinate, so that for any field $x$ one has $x(t,s=0)=x(t,s=2\pi)$. This is analogous to the so-called ``box quantization'' procedure employed in the canonical approach to Quantum Field Theory.\\
To deal with the gauge invariances it is useful to look at the BRST formulation of the model \cite{Govaerts:1991}; after the introduction of ghost degree of freedom and the BRST charge, the BRST extended constraints are readily calculated and, as expected from the properties of the duality, it is easy to check that their algebra is first class. In particular the (smeared) Poisson brackets read:
\begin{equation}
\begin{array}{r l}
\{ L^{\pm} (f) , L^{\pm} (g) \} =& \pm L^{\pm} (g_{,s} f- f_{,s} g)\ ,\\
\{ L^{\pm} (f) , L^{\emptyset} (g) \} =& 0\ , \\
\{L^{\pm}(f),L^{\mp}(g)\}=&-2\q \left (e^{Y/\xi}P_{A_1}L^{\emptyset}\right )\left (fg\right ) \ ,
\end{array}
\end{equation}
where $\pm$ label the two Virasoro generators, $\emptyset$ indicates the constraint related to the $U(1)$ gauge invariance and $P_{A_1}$ is the conjugate momentum of $A_1$. As expected these brackets are vanishing on the constraints surface. In the following we will choose to work in the conformal gauge for the gravitational sector, $\lambda_0=\lambda_1=1$, and the Coulomb gauge in the $U(1)$ sector, $A_0=A_{1,s}=0$. Gauge fixing is implemented by a suitable choice of the gauge fixing fermionic function.\\
Per effect of the decoupling, the BRST extended Virasoro generators are a sum of terms from the different sectors of the model:
\begin{equation}\label{BRSTgenerators}
L^{\pm} =L^{\pm, Z} + L^{\pm, Y} + L^{\pm, g} + \sum^N_{n=1}L^{\pm, \phi}_n \ ,
\end{equation}
where the $g$ superscript indicates the contributions of the BRST ghosts. This feature is of paramount importance in the quantization procedure, since it allows to quantize each sector separately. The different contributions take the form:
\begin{subequations}\label{virasoro_contributions}
\begin{align}
L^{\pm,Z}=&-\frac{1}{4}\left(P_{Z}\mp Z_{,s}\right)^{2}\mp\xi\left(P_{Z}\mp Z_{,s}\right)_{s}+\Lambda e^{Z/\xi}\ ,\label{virasoro_Z}\\
L^{\pm,Y}=&\frac{1}{4}\left(P_{Y}\pm Y_{,s}\right)^{2}\mp\xi\left(P_{Y}\pm Y_{,s}\right)_{s}+\q \Lambda e^{Y/\xi}P_{A_1}^{2} \ ,\label{virasoro_Y}\\
L^{\pm,\phi}=&\frac{1}{4}\left(P_{\phi}\pm \phi_{,s}\right)^{2}\ , \\
L^{\pm,g} = &\frac{\rmi}{2\pi}\left ( c^{\pm}b_{\pm,s}+2c^{\pm}_{,s} b_{\pm} \right )
\end{align}
\end{subequations}
where $P$ are the conjugate momenta of the fields and $c^\pm,b_\pm$ are the ghost fields associated with the Virasoro generators, satisfying:
\begin{equation}\label{ghostpsb}
\{c^\pm(s),b_\pm(s')\}^+=-2\rmi\pi\delta_{2\pi}(s -s')\ ,
\end{equation}
with $\delta_{2\pi}(s-s')$ being the $2\pi$-periodic Dirac $\delta$ distribution on the unit circle. Gauss' constraint on the other hand has no BRST extension and is simply given by $L^\emptyset=P_{A_1,s}$.
\section{Quantum theory}
\label{sec:quantization}
Quantization follows the usual canonical procedure, by choosing a polarization for phase space and a Hilbert space, by promoting observables to operators (using normal ordering when required, with the annihilation operators to the right of the creation ones) and by substituting Poisson brackets with commutators or anticommutators, as the case may be, inclusive of the extra factor $\rmi$ multiplying the values of the corresponding classical brackets.
\subsection{The central charge of the Virasoro algebra}
We will require the classical symmetries to be unbroken at the quantum level: the quantum operators corresponding to the $L$'s will have to exhibit a closed algebra. In particular we will make sure that no central term appears in the commutators.\\
The ${L}^\emptyset$ constraint has identically vanishing brackets with the two Virasoro generators and itself, hence the corresponding operator is quantum mechanically trivial. The Virasoro algebra, on the other hand will exhibit the typical central extension in each different sector $x$. Looking at the Fourier modes we will have:
\begin{equation}
\left[L_{r}^{\pm,x},L_{q}^{\pm,x}\right]= (r-q)\HB L_{r+q}^{\pm,x}+\HB^2 c_x\delta_{r+q} \ ,
\end{equation}
where $\HB$ is a place holder for $\hbar=1$. The ghosts and scalar matter degrees of freedom reproduce the usual results known from string theory \cite{Polchinski:1998rq, Green:1987sp}:
\begin{equation}
c_g=  -\frac{13}{6}r^{3}+\frac{1}{6}r\ , \qquad c_\phi = \frac{1}{12}  r^3 + \frac{1}{6}r \ ,
\end{equation}
In the Liouville sector, in order to ensure the closure of the quantum algebra, the possibility of quantum corrections to the coupling constant $\xi$ needs to be considered \cite{Curtright:1982}, in a manner dependent on the fields. As a matter of fact, only terms of \eqref{virasoro_Z} \eqref{virasoro_Y} involving the fields linearly need to be corrected, with the replacements $\xi \rightarrow \xi_Z = \xi + \delta_Z$ and $\xi \rightarrow \xi_Y = \xi + \delta_Y$ for the corresponding couplings, respectively. The factor $\xi$ appearing in the exponential Liouville term contributions to \eqref{virasoro_Z} \eqref{virasoro_Y} remains unchanged.
The commutators can be calculated explicitly by using expansions of the fields in terms of creation and annihilation operators and by fixing the value of the quantum corrections we can get a Virasoro algebra with a central extension. By choosing $\xi_Z =\xi_Y = \xi - \frac{\HB}{8 \pi \xi}$ we obtain the central charges:
\begin{subequations}
\begin{align}
c_Z= & - \left ( \frac{1}{12} + \frac{4 \pi}{\HB} \left ( \xi - \frac{\HB}{8 \pi \xi} \right )^2 \right ) r^3 - \frac{1}{6}r \ ,\\
c_Y = & \left ( \frac{1}{12} + \frac{4 \pi}{\HB} \left ( \xi - \frac{\HB}{8 \pi \xi} \right )^2 \right ) r^3 +\frac{1}{6}r\ ,
\end{align}
\end{subequations}
Note how the central charges in the $Z$ and $Y$ sector are identical in absolute value but appear with opposite signs. When summing up all contributions to the central charge the $c_Z$ and $c_Y$ cancel out and the only contributions are then given by the ghosts and the scalars\footnote{It is interesting to note that such a cancellation occurs only in the presence of a Maxwell field. As discussed in \cite{Govaerts:2011p3916,Zonetti:2012fk}, if the Maxwell field is not present at the classical level there is no Liouville potential of the $Y$ field. Therefore there is no need to fix a quantum correction to the coupling constant in the $Y$ sector, while quantum corrections to $\xi$ are still allowed. This freedom can be exploited to eliminate the cubic part of the central charge, determining also an upper bound on the number of scalar matter fields $N$. In this way the central charges for $Z$ and $Y$ cancel only partially and one is left with non vanishing vacuum fluctuations of the gravitational degrees of freedom.}:
\begin{equation}
c=\frac{N-26}{12}r^3 +\frac{N+1}{6} r \ .
\end{equation}
The central charge $c$ breaks the Virasoro algebra at the quantum level. However, it is possible to eliminate the $r^3$ term in $c$ by tuning the number of scalars to $N=26$, in the same way as the number of space-time dimensions is tuned in bosonic string theory. In this way the cubic term of the central charge is cancelled and the remaining linear term can be reabsorbed with a shift of the zero modes $L^\pm_0 \to L^\pm_0 -\HB 9/4$ so that the quantum Virasoro algebra is finally closed.\\
Such a shift of the zero modes is nothing else than the contributions of the quantum vacuum fluctuations to the total energy of the system, which are finite in two dimensions. 
\subsection{Quantum constraints}
Once the quantum Virasoro algebra is obtained, it is possible to find the quantum realization of the constraints on Hilbert space, following the usual Dirac prescription that physical states have to be annihilated by the constraints.
As a matter of fact the cosmological constant $\Lambda$ and the coupling constants $\xi$ and $\q$ are still free parameters: by requiring certain quantum states to be physical $\Lambda$ can be constrained to take a specific value.\\
The presence of the Liouville potentials involving the $Z$ and the $Y$ fields prevents one from following the standard string theory approach, {\it i.e.}, extracting the modes $L_n^\pm$ of the quantum constraints with a discrete Fourier transform and looking at $L^\pm_n |\psi_{phys} \rangle=0$ with $n = 0,1,2$. As discussed in Section \ref{sec:ccinqg} however, it is sufficient to use the weaker condition:
\begin{equation}\label{quantumconstraints}
\langle \psi_{phys} | L^\pm(s) | \psi_{phys} \rangle = 0,
\end{equation}
under the hypothesis/requirement that $|\psi_{phys}\rangle$ is physical.\\ This is because we are not looking to determine the set of physical states of the model, but rather to determine how the cosmological constant is constrained at the quantum level by the requirement that a given quantum state of the Universe is physical.\\
The space-coordinate dependence will have to be carried through and in some cases traded for a mode expansion via a Fourier transformation once the matrix elements between suitable states spanning the Hilbert space have been calculated. In particular, considering linear combinations of the shifted Virasoro generators, the quantum constraints for an arbitrary quantum physical state will be:
\begin{equation}
\langle L^+ + L^- \rangle  =0\ , \qquad \langle L^+ - L^- \rangle = 0\ , \qquad \langle L^\emptyset \rangle = 0 \ ,
\end{equation}
with:
\begin{equation}\label{Lshifted}
L^{\pm} = L^{\pm,Z} + L^{\pm,Y} + \sum_{n=1}^{N} L^{\pm,\phi_n} - \delta_m/2 \ ,
\end{equation}
inclusive of the quantum corrected coupling constants $\xi_Z=\xi_Y$ and the shift of the zero modes determined above. The ghost sector is omitted by taking advantage of the BRST invariance. As the cosmological constant enters the expressions for $ L^{\pm}$ only through the Liouville potential terms (cf. \eqref{virasoro_Z} \eqref{virasoro_Y}), which are identical for the $+$ and $-$ cases, only $\langle L^+ + L^- \rangle$ will depend on $\Lambda$. We have:
\begin{subequations}\label{qconstraints}
\begin{align}
\begin{split}
\langle L^+ + L^- \rangle =& -\frac{1}{2}\langle P_{Z}^2+ Z^2_{,s}\rangle +  \frac{1}{2}\langle P_{Y}^2+ Y_{,s}^2\rangle+ \sum_{n=1}^N \frac{1}{2}\langle P_{n}^2+ \phi_{n,s}^2\rangle +2 \xi\left ( \langle Z_{,ss} \rangle -\langle Y_{,ss} \rangle\right ) +\\&-\frac{\HB}{4\pi \xi}\left ( \langle Z_{,ss} \rangle -\langle Y_{,ss} \rangle\right )  - C_m +\Lambda \left [  2 \langle e^{Z/\xi} \rangle + 2 \q \langle e^{Y/\xi}P_{A_1}^{2}\rangle \right ],\end{split} \label{qconstraintcc}\\
\langle L^+- L^- \rangle =&  \langle P_{Z}Z_{,s}\rangle+\langle P_{Y}Y_{,s}\rangle + \sum_{n=1}^N \langle P_{\phi_n}\phi_{n,s}\rangle-2\left (\xi -  \frac{\HB}{8\pi \xi}\right )\left (\langle P_{Z,s}\rangle-\langle P_{Y,s}\rangle \right )  \ , \label{qconstraintm}\\
\langle L^\emptyset \rangle =& \langle P_{A_1,s}\rangle \ , \label{qconstraintg}
\end{align}
\end{subequations}
where $C_m = \HB 9/2$. A first qualitative analysis of the quantum constraints reveals interesting features. By solving \ref{qconstraints} for the cosmological constant we can see that contributions to $\Lambda$ are of three types (up to the factor in square brackets): classical contributions (the first line in \eqref{qconstraintcc}), quantum fluctuations of the vacuum (the $C_m$ term) and additional quantum contributions determined by the quantum corrections to the coupling constant (the first term in the second line of \eqref{qconstraintcc}). Schematically we can write the cosmological constant as:
\begin{equation}\label{schematic_cc}
\Lambda = -\left [ \left \langle \text{Kin}_{grav} \right \rangle + \left \langle \text{Kin}_{\phi} \right \rangle +2\left (\xi -  \frac{\HB}{8\pi \xi}\right ) \left \langle \partial^2_{grav} \right \rangle -  \delta_m \right ] \left \langle \text{Liouville} \right \rangle^{-1} \ .
\end{equation}
The QFT approach discussed in the introduction is recovered if all fields are taken to be in the vacuum state, so that only vacuum fluctuations of ghosts and scalars (finite in two dimensions when the regularization is removed) are left in the square brackets and the Liouville potential factor is unity\footnote{If quantum fluctuations of the gravitational sector are present, as in the case with no Maxwell field, an extra contribution $C_{grav}$ would appear in the square brackets.}.\\
It is more interesting, however, to look at the cosmological constant for a generic state away from the vacuum, when all terms can contribute to the value of $\Lambda$. We can expect the Liouville potentials, which are exponential in the fields and the inverse of the coupling constant $\xi$, to be very sensitive to excitations of the gravitational sector and/or small values of $\xi$ itself. For instance in the case of small $\xi$ we have $\langle \text{Liouville} \rangle \gg 0$, suppressing all terms in square brackets in \eqref{schematic_cc}. Let us choose $\xi=0.0035$ and let us require to be physical a quantum state in which all fields but $Z$ are in the vacuum and  $\langle Z \rangle = 1$. Then \eqref{qconstraintm} and \eqref{qconstraintg} vanish identically and \eqref{qconstraintcc} gives:
\begin{equation}
\Lambda \sim 1.85 \times 10^{-124}\ .
\end{equation}
This quantum state is nothing else than the vacuum used in the QFT approach supplemented by an excitation of the one gravitational degree of freedom in \eqref{dil-max_action}  (see \cite{Zonetti:2011ky} for details).
A physical interpretation of $\xi$ can be given from the condition \eqref{Ucondition} for the potential $U(X)$. Going a step back, the dilaton field $X$ in two-dimensional dilaton gravity can be seen, generally speaking, as the field that encodes the collective behaviour of the degrees of freedom that have been integrated out in the reduction process from higher dimensions: \emph{e.g.} in spherically reduced gravity $X$ is what is left of the angular degrees of freedom. On the other hand the single physical degree of freedom contained in the two-dimensional metric can always be taken to be a conformal mode, rescaling a Minkowski metric. From \eqref{Ucondition} we can see that $\xi$ regulates the relative weight of the kinetic term for the dilaton in the dilaton gravity action \eqref{dil-max_action}. A larger value for $\xi$ corresponds to a more ``frozen'' dynamics for $X$ w.r.t. the dynamics of the two-dimensional metric, while for small $\xi$ the kinetic term for $X$ dominates the action. In the same way, from \eqref{Gcondition}, we can see that a smaller $\xi$ corresponds to a stronger coupling of the dilaton with the Maxwell sector. Explicitly, for $\xi \ll 1$, we can rescale $X\to \xi X$ and $\q\to \q \xi^2$ and the classical action \eqref{dil-max_action} can be approximated as:
\begin{equation}\label{dil-max_small_xi}
 S_{\xi \ll 1}=\int_\mc{M} dx^{2}\sqrt{-g}\left(-\frac{1}{2}X_{,\mu}X^{,\mu}-2\Lambda-\frac{e^{X/\xi}}{16\q\Lambda}F_{\mu\nu}F^{\mu\nu}+O(\xi)\right)\ ,
\end{equation}
which to the zeroth order is a massless scalar non-linearly coupled to a Maxwell field on a conformally flat background. In our cosmological observations this seems a good approximation: at large scales the Universe is well described by a Minkowski metric and considering the short time scale during which observations are made we can approximate the conformal mode/scale factor as a constant. The remaining degrees of freedom then dominate the dynamics, justifying a choice of $\xi \ll 1$.\\
On the other hand for small $\xi$ an important contribution to $\Lambda$ can be given by the second of the $\left \langle \partial^2_{grav} \right \rangle$ terms, generated by the quantum corrections to the coupling constant. This suggests that quantum gravity contributions to the cosmological constant can play a fundamental role. If we take a closer look at the specific form of the contributions to $\Lambda$ in \eqref{qconstraints} we notice that interestingly the $Z$ and $Y$ field appear with opposite signs in all non-exponential terms, so that $\left \langle \text{Kin}_{grav} \right \rangle$, $\left \langle \partial^2_{grav} \right \rangle$ can have both positive or negative, allowing for cancellations of the $C_m$ contributions to $\Lambda$. Let us also point out that the Maxwell field $A_1$ exhibits a continuous spectrum in this model.\smallbreak
If we take the leap of thinking of the models treated here as effective descriptions of highly symmetrical four-dimensional theories we can consider the possibility that similar features may appear, including quantum gravity contributions to $\Lambda$ and excitations of the gravitational degrees of freedom that would renormalize the value of the cosmological constant in the fashion of  $\langle \text{Liouville} \rangle \gg 0$.
\subsection{Representation of operators and Hilbert space}
To have an explicit formulation of the schematic contributions to $\Lambda$ summarized in \eqref{schematic_cc} we need to choose how to represent operators and to describe quantum states. As a basis for Hilbert space two possibilities are at hand: coherent states, being eigenstates of the annihilation operators, have the advantage of providing rather simple expressions for the quantum constraints, and therefore seem to be the most obvious choice. Canonical coherent states however are not stable under the action of the quantum Hamiltonian, due to the presence of the Liouville potentials. Furthermore if we wish to look at the values for $\Lambda$ which follow from the requirement for the excitations of the spectrum of the theory to be physical, a Fock basis is the best option.\\
However if the quantum constraints are expressed in terms of creation and annihilation operators the exponential terms of the Liouville potentials in $L^{\pm}$ would spread every Fock excitation of the fields $Z,Y$ over the entire spectrum. To avoid this it is possible to use a diagonal representation for the constraint operators in the coherent state basis \cite{Klauder:1968dq}. This has the advantage of turning all the matrix elements calculations into Gaussian integrals over complex variables. By writing a general state as a tensor product of linear combinations of Fock excitations we will be able to obtain two constraint equations involving the cosmological constant $\Lambda$\footnote{We refer to \cite{Govaerts:2011p3916,Zonetti:2012fk} for details.}:
\begin{equation} \label{kernelrepconstraints}
L^\pm =  \int \prod_m \left[ \frac{d z_m d \bar{z}_m}{2 \pi} \right]  |\ul{z} \rangle \Biggl ( \Lslash^{\pm,Z} (s, z, \bar{z}) + \Lslash^{\pm,Y} (s, z, \bar{z}) + \sum_{n=1}^{N} \Lslash^{\pm,\phi_n} (s, z, \bar{z}) -C_m/2 \Biggr )\langle \ul{z} |\ ,
\end{equation}
where the $\Lslash$'s are suitable defined kernels and $m$ runs over all the modes of creation and annihilation operators. Coherent states are defined as:
\begin{equation}\label{coherentstate}
 | \ul{z} \rangle = \bigotimes_{f}^{ fields}\left ( \bigotimes_n |z_n^f \rangle \right )\ ,
\end{equation} 
where the first tensor product is over all fields excluding the ghosts.\\
Because of the decoupling the Hilbert space is a direct product of the Hilbert spaces for each field. Furthermore each field is described by a mode expansion, so that its Hilbert space is itself a tensor product of independent Hilbert spaces, one for each mode number $n$. For a generic field $f$ a completely general state may be written as:
\begin{equation}\label{physicalstatef}
| \psi^f(d) \rangle = \bigotimes_{n \in \mathbb{Z}} \left [ \sum_{\mu \geq 0} d_\mu^f(n) | \mu_n^f \rangle \right ]\ ,
\end{equation}
where $n$ labels the modes, $\mu^f_n$ is the occupation number of the mode $n$ of the field $f$, and the $d$'s are complex coefficients. 
Considering then the complete set of fields in the model, any state in the complete Hilbert space (inclusive of the quantum fields $Z,Y$ and $N$ free scalar fields $\phi_i$) can be written then as a sum of factorized states in the form of \eqref{physicalstatef}:
\begin{equation}\label{physicalstate}
|\psi\rangle = \sum_{\{d^Z, d^Y, d^n\}} | \psi^Z(d^Z) \rangle \otimes | \psi^Y(d^Y) \rangle \bigotimes_{n=1}^{N}| \psi^{\phi_n}(d^{\phi_n}) \rangle \ ,
\end{equation}
where the sum is over an arbitrary number of sets of $d$ coefficients.\\
This choice is not the most intuitive but it has the advantage of providing us with complete control on the single coefficients of each field, so that specific quantum states are easily selected for the purpose of a spectrum analysis\footnote{In the simple example of two decoupled systems, $A$ and $B$, with an Hilbert space basis $|n\rangle$ and $|m\rangle$ respectively, the easiest way to write a general state has the form $|\psi_g\rangle = \sum_{n,m} \psi(n,m)|n\rangle|m\rangle$. Considering factorized states $|\psi_f (a,b)\rangle = \sum_{n} a(n)|n\rangle\otimes \sum_mb(m)|m\rangle$, a sum over different sets of coefficients $\{a,b\}$ reproduces the general state given the identification $\psi(n,m) = \sum_{a,b} a(n)b(m)$, as in a series expansion.}.\\
To simplify the picture, and take advantage of the decoupling, without loosing insight in the mechanism that constrains the cosmological constant, we will consider a subset of Hilbert space, in which the quantum states \eqref{physicalstate} are defined with a single set of $d$ coefficients, so that the sum is dropped. In this way the quantum state is completely factorized, and we can work in each sector separately. The matrix elements for \eqref{kernelrepconstraints} will be in the form:
\begin{equation}\label{factorizedintegral}
 \langle L^\pm \rangle = \sum^{\text{\tiny fields}}_f\int \prod_m \left[ \frac{d z_m d \bar{z}_m}{2 \pi} \right] \Lslash^{\pm,f} (s, z, \bar{z})  | \psi^f (\ul{z}) |^2\ ,
\end{equation}
The explicit result of this calculation is reported in Appendix \ref{sec:matrixelements}. 
\subsection{An example of the spectrum of the cosmological constant}
To have an example of the spectrum of values of the cosmological constant that follow from the enforcement of the quantum constraints, we can look at an even simpler subset of the states \eqref{physicalstate}. In particular we can consider pure quantum states for the fields, \emph{i.e.} for each individual field $f$ and in each mode $n$ there is only a non vanishing $d_\mu^f(n)=1$ coefficient, so that \eqref{physicalstatef} reduces to:
\begin{equation}\label{pureex}
| \psi^f(d) \rangle = \bigotimes_{n \in \mathbb{Z}} | \bar{\mu}_n^f \rangle\ ,
\end{equation}
and the constraints read:
\begin{equation}
\Lambda=\left(\prod_{\ell} \T_{\ell}^{Z}+qP^{2}\prod_{\ell} \T_{\ell}^{Y}\right)^{-1}\left[\hbar\frac{9}{4}+\frac{1}{4\pi}\sum_{f}^{\mbox{\tiny fields}}\beta\left(f\right)\Bigl(\sum'_{n}+2\delta_{0}^{n}\Bigr)\left(n^{*}\frac{\left(\bar{\mu}^{f}+1\right)!-1}{\bar{\mu}^{f}!}\right)\right]\ ,
\end{equation}
\begin{equation}
\sum_{f}^{\mbox{\tiny fields}}\beta\left(f\right)\sum'_{n}n\bar{\mu}^{f}=0\ ,
\end{equation}
with:
\begin{equation}
\T_{\ell}^{f}=\sum_{m=0}^{\mu^{f}}\left(\frac{\mu^{f}!}{(\mu^{f}-m)!}\right)^{2}\frac{1}{m!}\left(\frac{1}{4\pi\xi^{2}}\right)^{\mu^{f}-m}\ ,
\end{equation}
and:
\begin{equation}
n^* = \Biggl \{ \begin{array}{c c} \frac{1}{4}&\ : n = 0\ ,\\ |n|&\ : n\neq 0\ ,\end{array}\qquad 
\beta(f) = \Biggl \{ \begin{array}{c c} -1& : f = Z\ ,\\ 1 & : f = Y,\phi\ ,\end{array} \qquad  P^2=\langle P_{A_1} \rangle \ .
\end{equation}
Note that in this case there is no contribution proportional to the quantum corrected coupling constants $\xi_Z=\xi_Y$, while the classical $\xi$ appears in $\T_\ell^f$. Most importantly, there is no dependence of the constraints on the space coordinate $s$. If we take all $\bar{\mu}^f=0$ we reduce ourselves to the vacuum, and one has simply $\Lambda = \HB 9/4$.
\begin{figure}
\captionsetup[subfigure]{labelformat=empty}
\subfloat[]{\label{fig:pos_cc_spectrum} \includegraphics[width=.45\textwidth]{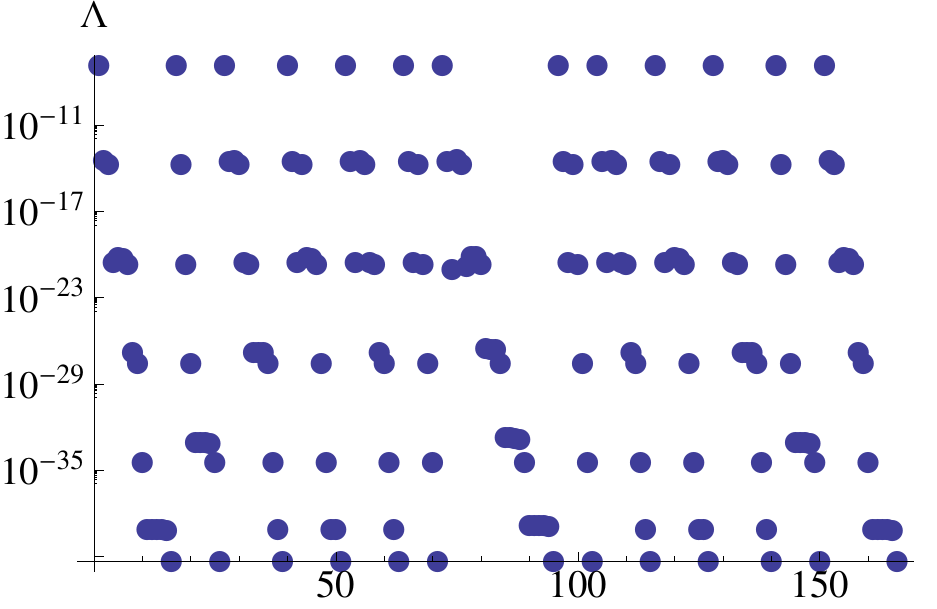}}
\subfloat[]{\label{fig:neg_cc_spectrum} \includegraphics[width=.45\textwidth]{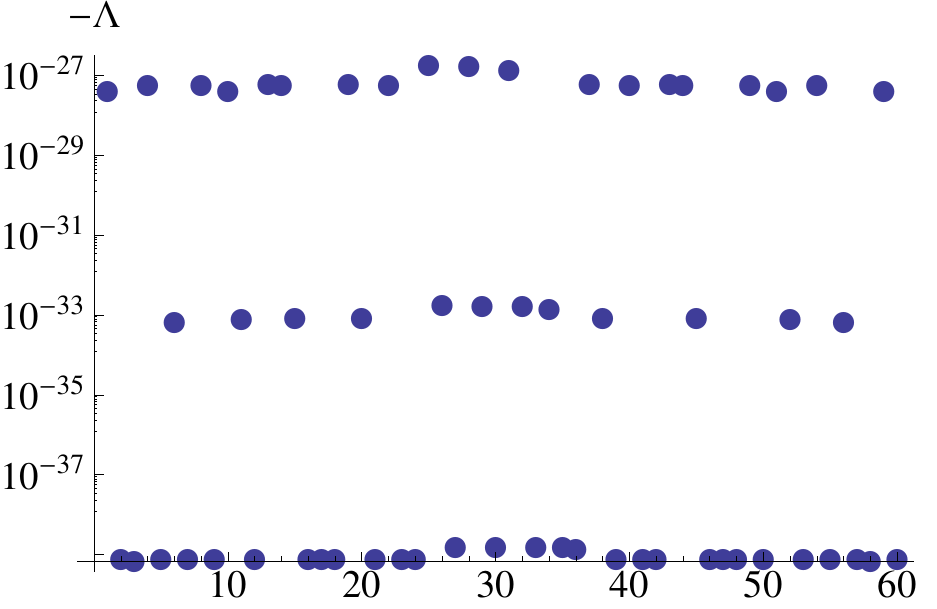}}
\caption{Positive (left) and negative (right) values of the cosmological constant lie on discrete levels ranging, in absolute value, from $10^{-10}$ to $10^{-39}$ with the choices $\HB = 1,\ \xi = 10^{-4},\ qP^2=1$. The horizontal axis simply labels sequentially the values.}
\label{fig:cc_spectrum}
\end{figure}
The second constraint restricts the excitations of the fields: considering that $n\in \mathbb{Z}$ and $\bar{\mu}\in \mathbb{Z^+}$ many configurations will fail to satisfy the constraint.\\ It is possible to visualize the allowed values for $\Lambda$ in a further restriction of \eqref{pureex}: we can take all matter fields in the vacuum, $\bar{\mu}^\phi=0$, and consider the $Z$ and $Y$ fields to be excited in a single mode only:\begin{equation}
|\psi^Z\rangle = |\mu_{Z}\rangle\bigotimes_{n\neq n^Z} |0\rangle, \quad|\psi^Y\rangle = |\mu_{Y}\rangle\bigotimes_{n\neq n^Y} |0\rangle,\quad |\psi^\phi\rangle = \bigotimes_{n} |0\rangle,
\end{equation}
We can then solve the constraints for all combinations of:
\begin{equation}\label{n_bounds}
\mu_{Z},\mu_{Y}\in [1,10] \qquad n_Z,n_Y \in [-10,10]\ ,
\end{equation}
and take, for instance, $\HB = 1,\ \xi = 10^{-4},\ qP^2=1$. We can then plot the list of values for $\Lambda$, separately for positive and negative portions of the spectrum, in Fig. \ref{fig:cc_spectrum}.
As we see the values for the cosmological constant lie on a set of different discrete levels ranging from $10^{-10}$ to $10^{-37}$ and multiple quantum states correspond to each value.
Generally speaking we expect the same behaviour if the bounds \eqref{n_bounds} are lifted, with a countable infinity of quantum states corresponding to each of the infinitely many discrete values for the cosmological constant. The inclusion of mixed quantum states on the other hand introduces a number of real parameters (combinations of the complex $d$ coefficients in \eqref{physicalstatef}).
\section{Discussion}\label{sec:discussion}
In this work we discussed the cosmological constant problem in quantum gravity, with a particular focus on the case of two-dimensional models, intended as an effective description of highly symmetric higher dimensional models.\\
We have argued that in any (canonical) quantum theory of gravity the requirement of any quantum state to be physical provides, through the expectation values of the quantum constraints, a set of equations that can be solved for the cosmological constant, thus determining its value as a function of the quantum numbers labelling quantum states.
This general idea has been applied in detail to the case of a two-dimensional dilaton gravity coupled to a Maxwell and scalar matter fields, a class of models that under specific conditions can be quantized non-perturbatively in the canonical approach.\\
By requiring the expectation values of the quantum constraints to vanish for generic quantum states we were able to show that the value of the cosmological constant is determined not only by the vacuum fluctuations of the quantum fields (both in the matter and gravitational sectors, even if cancellations occur), but also by dynamical contributions, including quantum corrections to effective coupling constants, and a renormalization factor of gravitational origin.
In particular for small values of the effective coupling constant $\xi$, which corresponds to a frozen dynamics of the two-dimensional conformal factor w.r.t. the dynamics of the two-dimensional dilaton, the renormalization factor can be very small, resulting in a very small value for $\Lambda$ in natural units. The usual result of QFT of a cosmological constant proportional to the vacuum energy is recovered if the state required to be physical is the vacuum of the theory.\\
While the validity of this result is restricted to the specific framework of the two-dimensional dilaton gravity models considered (even though they can be motivated from higher dimensional considerations), it is a non-trivial example that quantum gravity can in fact contribute in a fundamental way to the value of the cosmological constant. In particular we cannot exclude the possibility of non-perturbative contributions in higher dimensions, such as the quantum corrections to the coupling constants, that can deeply change the picture.\\
At this point we are able to determine $\Lambda$ for any given quantum states. If we believe that this approach might in fact predict a meaningful value for the cosmological constant, we are now faced with the issue of determining which one is the quantum state the Universe is in.

\begin{acknowledgments}
SZ warmly thanks Daniel Grumiller for the many useful discussions and suggestions, as well as for the hospitality extended to him at the Institute of Theoretical Physics of TUWien. SZ benefits from a PhD research grant of the Institut Interuniversitaire des Sciences Nucl\'eaires (IISN, Belgium); this work is supported by the Belgian Federal Office for Scientific, Technical and Cultural Affairs through the Interuniversity Attraction Pole P6/11.
\end{acknowledgments}
 
\appendix

\section{Matrix elements of the quantum constraints}\label{sec:matrixelements}
The projection of \eqref{physicalstate} on the coherent states \eqref{coherentstate} gives, explicitly:
\begin{equation}
| \psi (\ul{z}) |^2 = \prod_{f \in \text{\tiny fields}}| \psi^f (\ul{z}) |^2= \prod_{f \in \text{\tiny fields}} \prod_n \left [\sum_{\mu, \nu \geq 0} d^f_\mu(n) \bar{d}^{f}_\nu (n) \bar{z}^\mu_n z^\nu_n e^{-|z_n|^2} \right ]\ .
\end{equation}
By virtue of the factorization the constraint equations \eqref{quantumconstraints} reduce to a sum of independent integrals over complex variables:
\begin{equation}
\begin{split}
 \langle L^\pm \rangle = &\int \prod_m \left[ \frac{d z_m d \bar{z}_m}{2 \pi} \right] \Lslash^{\pm,Z} (s, z, \bar{z})  | \psi^Z (\ul{z}) |^2+\int \prod_m \left[ \frac{d z_m d \bar{z}_m}{2 \pi} \right] \Lslash^{\pm,Y} (s, z, \bar{z})  | \psi^Y (\ul{z}) |^2 +\\  &+ \sum_{n=1}^{N} \int \prod_m \left[ \frac{d z_m d \bar{z}_m}{2 \pi} \right]\Lslash^{\pm,\phi_n} (s, z, \bar{z})| \psi^{\phi_n} (\ul{z}) |^2- \frac{\delta_m}{2} .
\end{split}
\end{equation}
where the kernel of a generic operator $\mc{O}$ is calculated as in
\begin{equation}\label{kernel}
\Oslash (z, \bar{z}) =  \mbox{exp}\left( \sum_m\partial_{z_m} \partial_{\bar{z}_m} \right) \langle \ul{z} |\mc{O} |\ul{z} \rangle \ .
\end{equation}
The integrals are all Gaussian in the $z$'s, since $|\psi|^2$ carries a Gaussian factor for each mode. In a generalization of the formulas contained in \cite{Zonetti:2012fk} we can write the quantum constraints \eqref{qconstraints} as:
\begin{equation}
\begin{split}\label{qc1}
\langle L^{+}+L^{-}\rangle  = & \frac{2}{\sqrt{\pi}}\sum'_{n}|n|^{3/2}\left[\frac{\xi_{Y}}{\varpi_{n}^{Y}}\Im\left(\omega_{n}^{(1)Y}e^{-ins}\right)-\frac{\xi_{Z}}{\varpi_{n}^{Z}}\Im\left(\omega_{n}^{(1)Z}e^{-ins}\right)\right]-\hbar\frac{9}{2}-\\
&-\frac{1}{4\pi}\sum_{f}^{\mbox{\tiny fields}}\beta(f)\Biggl\{\Bigl(\sum_{n,m\geq0}+\sum_{n,m\leq0}\Bigr)_{n \neq m}\frac{4\sqrt{n^{*}m^{*}}}{\varpi_{n}^{f}\varpi_{m}^{f}}\Re\left(\omega_{n}^{(1)f}e^{-ins}\right)\Re\left(\omega_{m}^{(1)f}e^{-ims}\right)+\\
& +\Bigl(\sum'_{n}+2\delta_{0}^{n}\Bigr)\left[\frac{2n^{*}}{\varpi_{n}^{f}}\left(\Re\left(\omega_{n}^{(2)f} e^{-i2ns}\right)+\tilde{\omega}_{n}^{f}-1\right)\right]\Biggr\} +2\Lambda\left(\prod_{\ell}\T_{\ell}^{Z}+qP^{2}\prod_{\ell}\T_{\ell}^{Y}\right)\ ,
\end{split}
\end{equation}
\begin{equation}
\begin{split}\label{qc2}
\langle L^{+}-L^{-}\rangle =&\frac{2}{\sqrt{\pi}}\sum'_{n}n|n|^{1/2}\left[\frac{\xi_{Y}}{\varpi_{n}^{Y}}\Im\left(\omega_{n}^{(1)Y}e^{-ins}\right)-\frac{\xi_{Z}}{\varpi_{n}^{Z}}\Im\left(\omega_{n}^{(1)Z}e^{-ins}\right)\right]-\\ &-\frac{1}{4\pi}\sum_{f}^{\mbox{\tiny fields}}\beta\left(f\right)\Biggl\{\Bigl(\sum_{n,m\geq0}-\sum_{n,m\leq0}\Bigr)_{n\neq m}\frac{4\sqrt{n^{*}m^{*}}}{\varpi_{n}^{f}\varpi_{m}^{f}}\Re\left(\omega_{n}^{(1)f}e^{-ins}\right)\Re\left(\omega_{m}^{(1)f}e^{-ims}\right)+\\ &+\sum'_{n}\left[\frac{2n}{\varpi_{n}^{f}}\left(\Re\left(\omega_{n}^{(2)f}e^{-i2ns}\right)+\tilde{\omega}_{n}^{f}\right)\right]\Biggr\}\ ,
\end{split}
\end{equation}
where:
\begin{equation}
n^* = \Biggl \{ \begin{array}{c c} \frac{1}{4}&\ : n = 0\ ,\\ |n|&\ : n\neq 0\ ,\end{array}\qquad
\beta(f) = \Biggl \{ \begin{array}{c c} -1& : f = Z\ ,\\ 1&\ : f = Y,\phi\  ,\end{array}
\end{equation}
and
\begin{subequations}\label{omegas}
\begin{align}
\varpi^f_n =& \sum_{\mu \geq 0} |d^f_\mu(n) |^2 \mu!\ ,\\
\omega^{(1)f}_n = & \sum_{\mu \geq 0} d^f_\mu(n) \bar{d}^f_{\mu+1}(n)(\mu + 1)!\ ,\\
\omega^{(2)f}_n = & \sum_{\mu \geq 0} d^f_\mu(n) \bar{d}^f_{\mu+2}(n)(\mu + 2)!\ ,\\
\tilde{\omega}^f_n =& \sum_{\mu \geq 0} |d^f_\mu(n) |^2 (\mu+1)!\ ,
\end{align}
\end{subequations}
while
\begin{equation}
\T_{\ell}^{f}=\sum_{\mu,\nu\ge0}d_{\mu}^{f}(n)\bar{d}_{\nu}^{f}(n)\sum_{m=0}^{\nu}\frac{\nu!\mu!}{(\nu-m)!m!(\mu-m)!}\left(ie^{-i\ell s}\right)^{\mu-\nu}\left(2\xi\sqrt{\pi}\right)^{\mu+\nu-2m} \ .
\end{equation} 
 
\bibliographystyle{hunsrt}
\bibliography{references}

\end{document}